\documentclass[reprint,superscriptaddress,
amsmath,amssym,
aps,prb,longbibliography]{revtex4-2}
\usepackage[colorlinks=true, urlcolor=blue, linkcolor=blue, citecolor=blue]{hyperref}
\usepackage{graphicx}
\usepackage{dcolumn}
\usepackage{bm}
\usepackage[mathlines]{lineno}
\usepackage{xcolor}
\usepackage{float}

\begin{document}

\title{Broken site symmetry of Fe adatoms on Bi$_2$Te$_3$}

\author{Duy Nguyễn}
\affiliation{Department of Physics, \href{https://physics.osu.edu}{The Ohio State University}, Columbus, Ohio 43210, USA}

\author{Hari Paudyal}
\author{Joseph R. Sink}
\author{Michael E. Flatté}
\affiliation{Department of Physics and Astronomy, \href{https://physics.uiowa.edu}{University of Iowa}, Iowa City, Iowa 52242, USA}

\author{Jay A. Gupta}
\email[Contact author: ]{gupta.208@osu.edu} 
\affiliation{Department of Physics, \href{https://physics.osu.edu}{The Ohio State University}, Columbus, Ohio 43210, USA}
\begin{abstract}
We report a combined scanning tunneling microscopy and atomistic theoretical study of Fe adatoms on the Bi$_2$Te$_3$(111) surface. Topographic imaging at $4.5$~K shows that Fe adatoms in fcc and hcp hollow sites exhibit a threefold-symmetric contrast, consistent with the $C_{3v}$ symmetry of the adsorption site. However, simultaneously acquired differential conductance ($dI/dV$) maps reveal a pronounced reduction in symmetry, evidenced by differential contrast observed at nearest-neighbor Te sites. Density functional theory calculations show that the Fe/Bi$_2$Te$_3$ system undergoes a static Jahn--Teller distortion, reducing the adsorption symmetry from $C_{3v}$ to $C_{1v}$, with the distorted configuration favored by $72.5$~meV. Orbital-projected density of states calculations show that the occupied states near the Fermi level are dominated by $d_{xz}$ and $d_{yz}$ orbitals, whereas the unoccupied states are primarily of $d_{z^2}$, $d_{x^2-y^2}$, and $d_{xy}$ character. The local density of states from these orbitals is in good qualitative agreement with experimental $dI/dV$ spectra. Furthermore, simulated local-density-of-states maps using a tight-binding Green's function approach are consistent with experimental $dI/dV$ maps, confirming that the reduced symmetry originates from the $C_{1v}$ structural distortion.
\end{abstract}
\maketitle
\section{Introduction}
Three-dimensional topological insulators (3D TIs), such as Bi$_2$(Te, Se)$_3$ and Sb$_2$Te$_3$, host topological surface states (TSS) protected by time-reversal symmetry (TRS), which can be broken through the introduction of magnetic order~\cite{qi_topological_2008, liu_magnetic_2009, qi_topological_2011}. This interaction can open a magnetic gap~\cite{chen_massive_2010}, induce complex spin textures~\cite{laref_induced_2020}, and produce topological transport such as the quantum anomalous Hall effect~\cite{chang_experimental_2013, lee_imaging_2015}. These phenomena motivate efforts to understand, control, and spatially resolve magnetic order at TI surfaces. 

Previous studies have investigated the interaction of magnetic adatoms with TSS, introducing transition metals such as Mn, Co, and Fe onto the surfaces of Bi$_2$Te$_3$~\cite{sessi_signatures_2014, eelbo_strong_2014, west_identification_2012}. Prior work using scanning tunneling microscopy (STM) and X-ray magnetic circular dichroism (XMCD) identified fcc and hcp hollow-site adsorption geometries for Fe adatoms on the Te-terminated surface, and reported strong out-of-plane magnetic anisotropy for both sites~\cite{eelbo_strong_2014}. However, these results have not been connected with detailed local density of states (LDOS) imaging and spectroscopy at the adsorption site, and magnetic adatoms on Bi$_2$(Te, Se)$_3$ have been assumed to retain the high symmetry (C$_{3v}$) of the hollow site. 

Here, we report a combined STM and atomistic theoretical study of the adsorption geometry and electronic structure of Fe adatoms on the Te-terminated surface of Bi$_2$Te$_3$. STM topographic imaging reveals Fe adatoms occupy fcc and hcp hollow sites, with an apparent threefold symmetry expected for the site. However, for fcc-Fe, differential conductance ($dI/dV$) imaging and spectroscopy reveal a reduced symmetry not reported previously. Our density functional theory (DFT) calculations indicate this is due to a static Jahn-Teller (JT) distortion in which fcc-Fe adatoms relax into a C$_{1v}$ configuration rather than the commonly assumed C$_{3v}$ geometry. Tight-binding Green's function simulations incorporating this distortion reproduce the experimentally observed LDOS features, confirming the role of JT symmetry breaking. Our findings offer insight into how JT distortions may influence the magnetic anisotropy of transition-metal impurities on TI surfaces. 

\section{Methods}
All STM measurements were conducted at $4.5$~K in an ultrahigh vacuum (UHV) system ($10^{-11}$~mbar) using a CreaTec LT-STM system. All data presented here were taken with a Ni tip, though similar results were obtained with non-magnetic W and PtIr tips. Scanning tunneling spectra ($dI/dV$) were acquired via lock-in detection with an AC bias modulation at $877$~Hz and sweeping voltage with the feedback loop disabled. To emphasize Fe-related features, spectra were normalized to a (smoothed) $I(V)$ curve. Spatial $dI/dV$ maps were attained by recording the lock-in $dI/dV$ signal with the feedback on while the tip was scanned. STM images were processed with the Gwyddion software package~\cite{Necas2012}. The Bi$_2$Te$_3$ single crystal was purchased from 2D Semiconductors. To expose a clean surface, a small metal post was attached to the crystal surface using UHV-compatible EPO-TEK\textsuperscript{\tiny\textregistered} H21D epoxy, cured at $150^\circ$C for 1 hour. The sample was then transferred into the UHV chamber at room temperature, where the post was mechanically knocked off using a translator to expose a fresh Bi$_2$Te$_3$(111) surface. After cleaving, the sample was loaded into the STM and cooled to $4.5$~K. Fe was deposited from an in situ electron-beam evaporator while the sample was held at $4.5$~K. Prior to deposition, the Fe source was degassed, and evaporation at an e-beam power of $20$~W for $20$~s yielded a coverage of $\sim 0.2\%$ of a monolayer.

First-principles calculations were performed using the Vienna \textit{Ab initio} Simulation Package~\cite{furthmuller1996dimer, kresse1996efficiency} with projector augmented wave pseudopotentials~\cite{blochl1994projector} and a Perdew--Burke--Ernzerhof exchange-correlation functional~\cite{perdew1996o} within the generalized gradient approximation. A plane wave energy cutoff of 500~eV, a smearing value of 0.1~eV, and a $\Gamma$-centered $3\times3\times1$ Monkhorst-Pack \textbf{k}-mesh were used in structural relaxation and electronic structure calculations. The self-consistent energy convergence criterion of 10$^{-6}$~eV was used in the calculations of the atomic and electronic structures, whereas the atomic positions and lattice constants were relaxed until the maximum force on each atom was less than 10$^{-4}$~eV/\AA. Spin-orbit coupling effects were incorporated into self-consistent field calculations, with the magnetization oriented along the $c$-axis. 

The spatial dispersion of a defect state (i.e., the local density of states, $\eta\sim dI/dV$) is sensitive to the electronic properties in the vicinity. For this reason, we utilize a hybrid method for computing the Fe-adatom wave function that leverages DFT's ability to determine accurate ground state configurations and bonding chemistry, and a real-space tight-binding Green's function formalism that allows for the embedding of a single defect in an otherwise pristine/infinite 2D slab with an electronic dispersion and total energy scale closely matched to the experiment. The Fe-adatom is treated as a localized impurity potential that hybridizes the Fe $d$-orbitals with the surface nearest-neighbor Te $p$-orbitals. The $C_{1v}$ distortion identified by DFT was included by displacing the Fe atom from the high-symmetry $C_{3v}$ fcc hollow site toward a nearest-neighbor (nn-Te) atom.

\begin{figure}[t]
\centering
\includegraphics[width=0.51\textwidth]{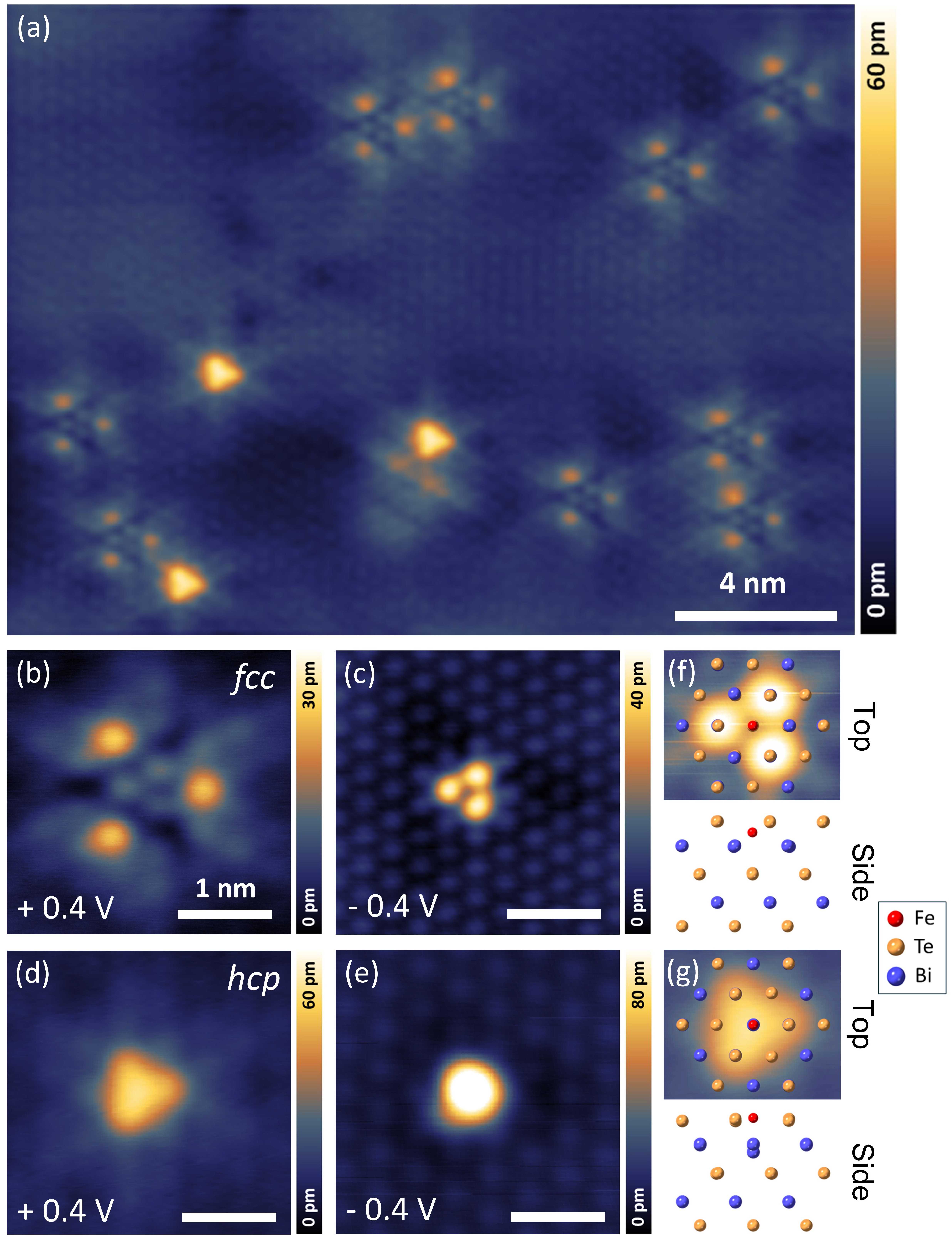}
\caption{
STM identification of fcc- and hcp-Fe adsorption sites on Bi$_2$Te$_3$.
(a) Large-area constant-current STM topograph of Fe adatoms deposited on the Te-terminated Bi$_2$Te$_3$(111) surface ($21$~nm$\times 16$~nm, $V = +0.4$~V, $I_{\text{set}} = 0.1$~nA), showing that Fe predominantly occupies fcc hollow sites.
(b, c) High-resolution images of an fcc-Fe adatom acquired at sample-bias voltages of $+0.4$~V and $-0.4$~V, respectively ($I_{\text{set}} = 0.3$~nA).
(d, e) Corresponding images of an hcp-site Fe adatom acquired under the same tunneling conditions.
(f, g) Top and side views of atomic models for Fe adsorbed at the fcc and hcp hollow sites, respectively. Fe, Te, and Bi atoms are shown in red, orange, and blue, respectively.}
\label{fig1}
\end{figure}

\section{Results and discussion}

Figure~\ref{fig1}(a) shows a low coverage of Fe adatoms imaged with two distinct, triangular-shaped contrasts. High resolution images indicate both contrasts are centered on hollow sites in the surface Te lattice. There are two flavors of hollow site with fcc-like and hcp-like stacking that are not distinguished in the STM images. The assignments of fcc-Fe and hcp-Fe adatoms in STM images are based on good agreement with DFT calculations, here and in prior work \cite{eelbo_strong_2014, west_identification_2012}. For fcc-Fe, the empty-state image at $+0.4$~V displays three bright outer lobes, while the filled-state image at $-0.4$~V shows the intensity shifted inward to three inner lobes. An overlay with the surface lattice in Fig.~\ref{fig1}(f) indicates that these inner and outer lobes are aligned with the nn-Te and next-nearest-neighbor-Te (nnn-Te) atoms surrounding the Fe atom. In contrast, the hcp-Fe adatoms show the strongest intensity at the nn-Te sites in both the empty and filled states, and exhibit a more isotropic, nearly circular contrast. We find that the fcc-site is more frequently occupied than the hcp-site, with an fcc:hcp occupancy ratio of approximately $3$:$2$, in good agreement with that in Ref.~\cite{eelbo_strong_2014}. We also find that hcp-Fe are readily nudged to the fcc-site by STM tip pulsing or repeated scanning, and we will focus our attention on fcc-Fe in the discussion below.

\begin{figure}[h]
\centering
\includegraphics[width=0.45\textwidth]{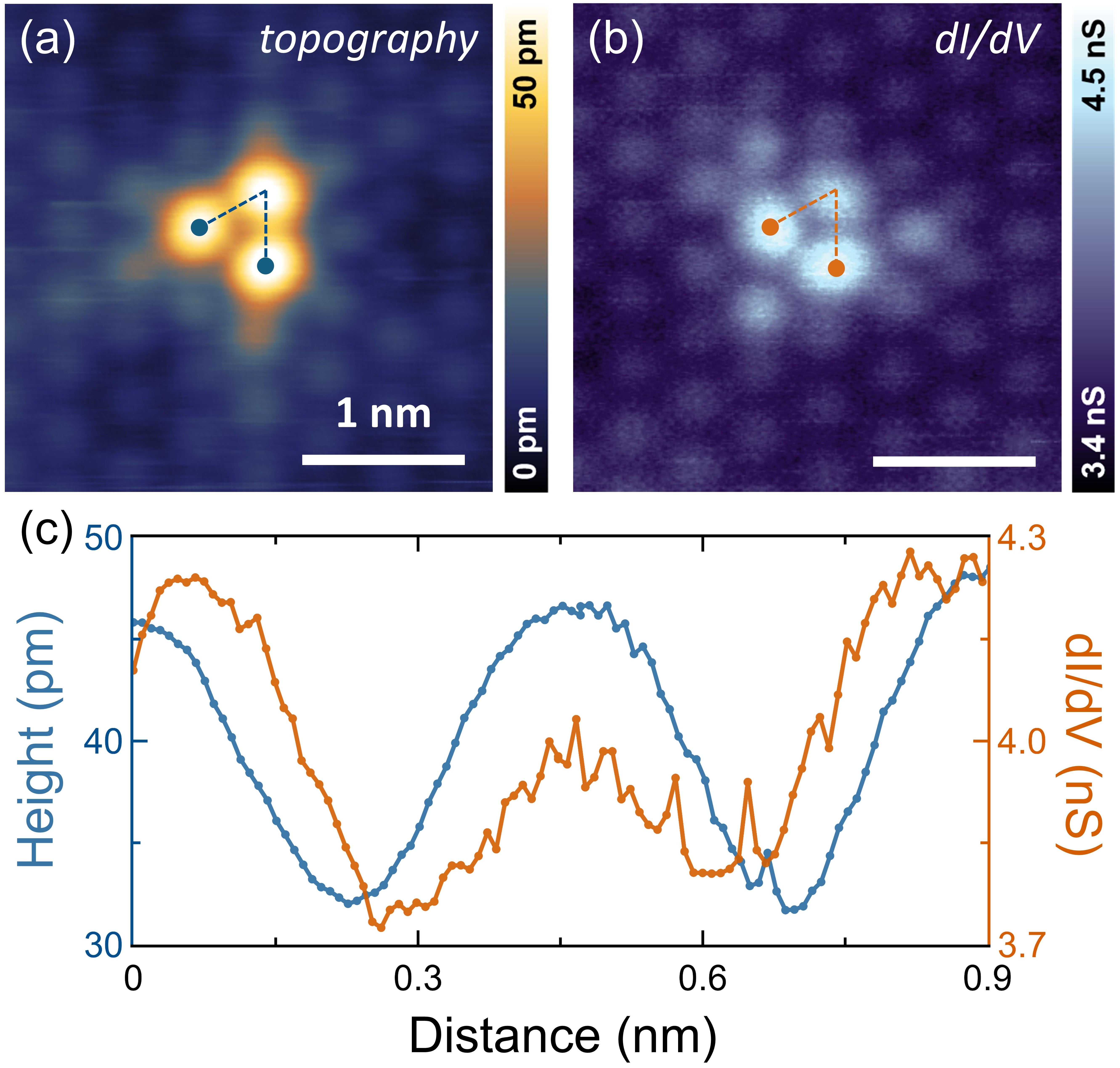}
\caption{
(a) Constant-current image of an fcc-site Fe shows three bright lobes exhibiting near threefold symmetry. 
(b) Simultaneously acquired $dI/dV$ image reveals unequal intensity among the lobes, indicating reduced symmetry ($V = -0.2$~V, $I_{\text{set}} = 0.3$~nA, $V_\text{mod} = 25$~mV).
(c) Line profiles extracted along the dashed path in (a) and (b) compare the topographic height (blue) and $dI/dV$ intensity (orange). The topography exhibits peak height variations by only $\sim 3$~pm between lobes, while showing larger changes in $dI/dV$.}
\label{fig2}
\end{figure}

\begin{figure}[hb]
\centering
\includegraphics[width=0.35\textwidth]{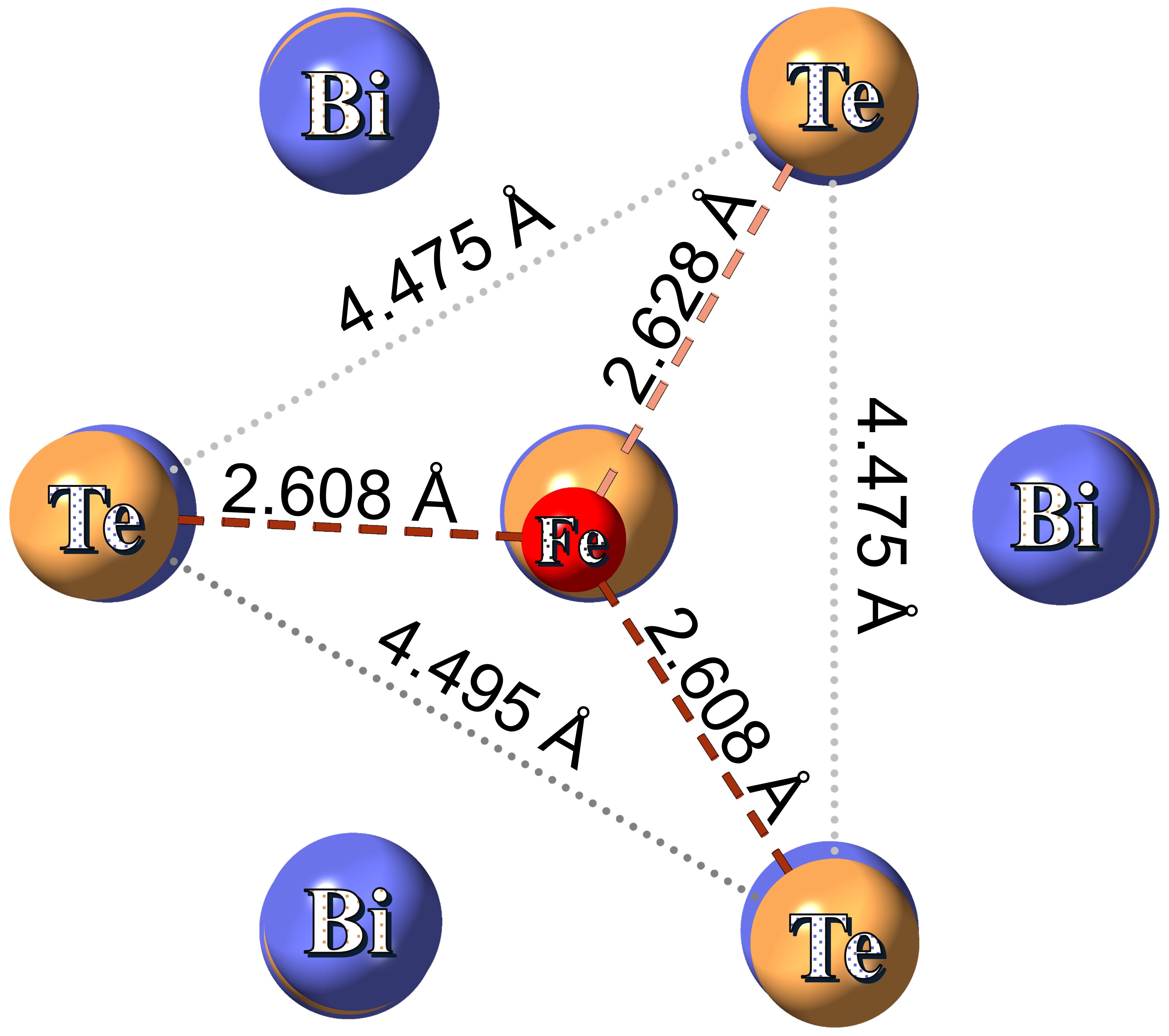}
\caption{Calculated structural relaxation of the fcc-site Fe adatom on Bi$_2$Te$_3$. Relaxed adsorption geometry showing displacement of the Fe atom from the symmetric $C_{3v}$ hollow site toward a nn-Te atom, reducing the symmetry to $C_{1v}$.}
\label{fig3}
\end{figure}

\begin{figure*}[ht!]
\centering
\includegraphics[width=.9\textwidth]{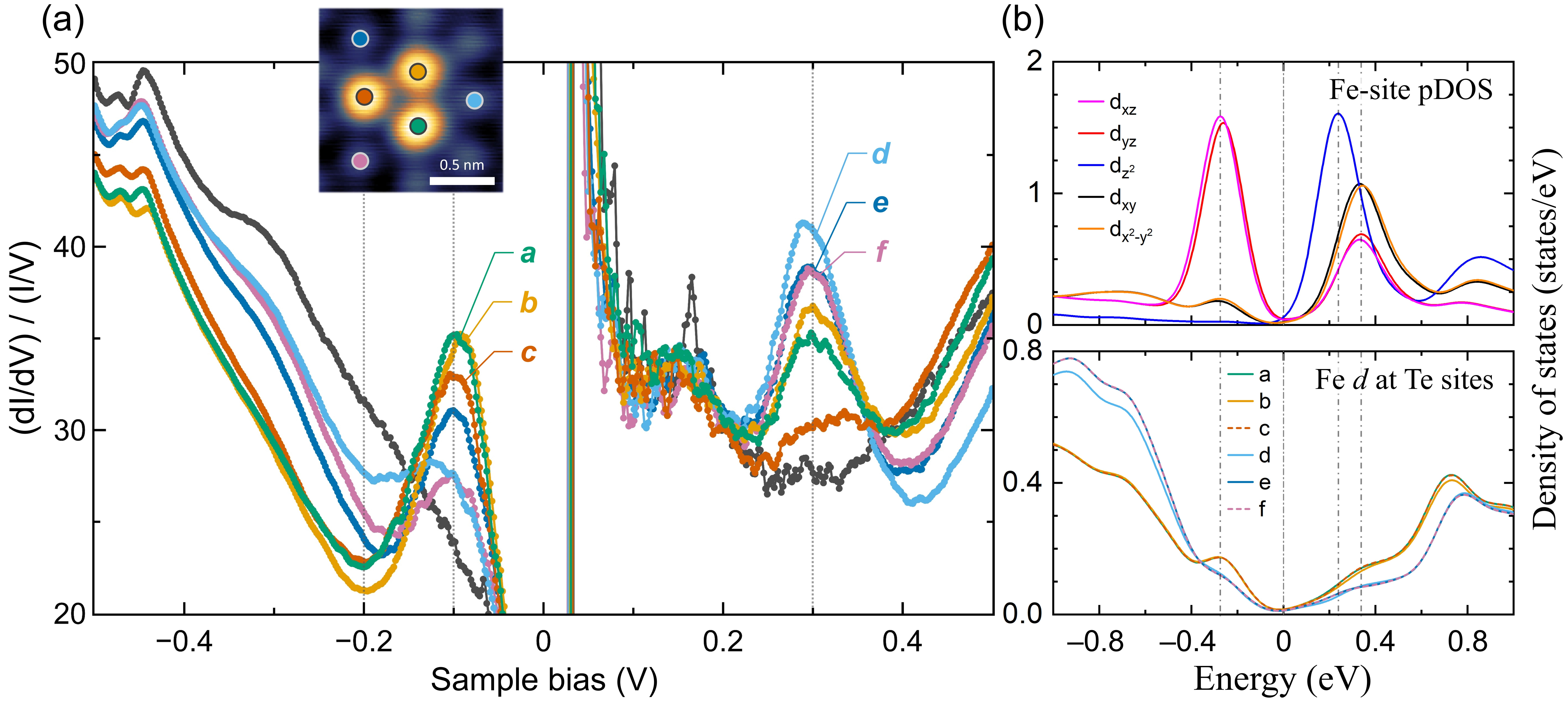}
\caption{
(a) Normalized $dI/dV$ spectra acquired on an fcc-site Fe adatom in Bi$_2$Te$_3$ reveal a common set of impurity resonances in the occupied and unoccupied states, with the spectral weight varying between the nn-Te and nnn-Te. The locations at which the spectra were acquired are indicated by correspondingly colored dots in the inset STM image; the Bi$_2$Te$_3$ spectrum is shown in gray. The spectra were acquired at a setpoint of $V_{\text{set}} = -0.5$~V and $I_{\text{set}} = 2$~nA.
(b) pDOS for the relaxed fcc-site Fe adatom configuration. Upper: Fe-site pDOS resolved by $d$-orbital character, showing pronounced peaks dominated by $d_{xz}$ and $d_{yz}$ in the occupied states and $d_{z^2}$, $d_{x^2-y^2}$and $d_{xy}$ in the unoccupied states. Lower: Fe $d$-orbital weight projected at the six neighboring Te sites, corresponding to the spectra positions in (a). Traces b and c (nn-Te) and traces e and f (nnn-Te) overlap.}
\label{fig4}
\end{figure*}

\begin{figure*}[ht!]
\centering
\includegraphics[width=0.9\textwidth]{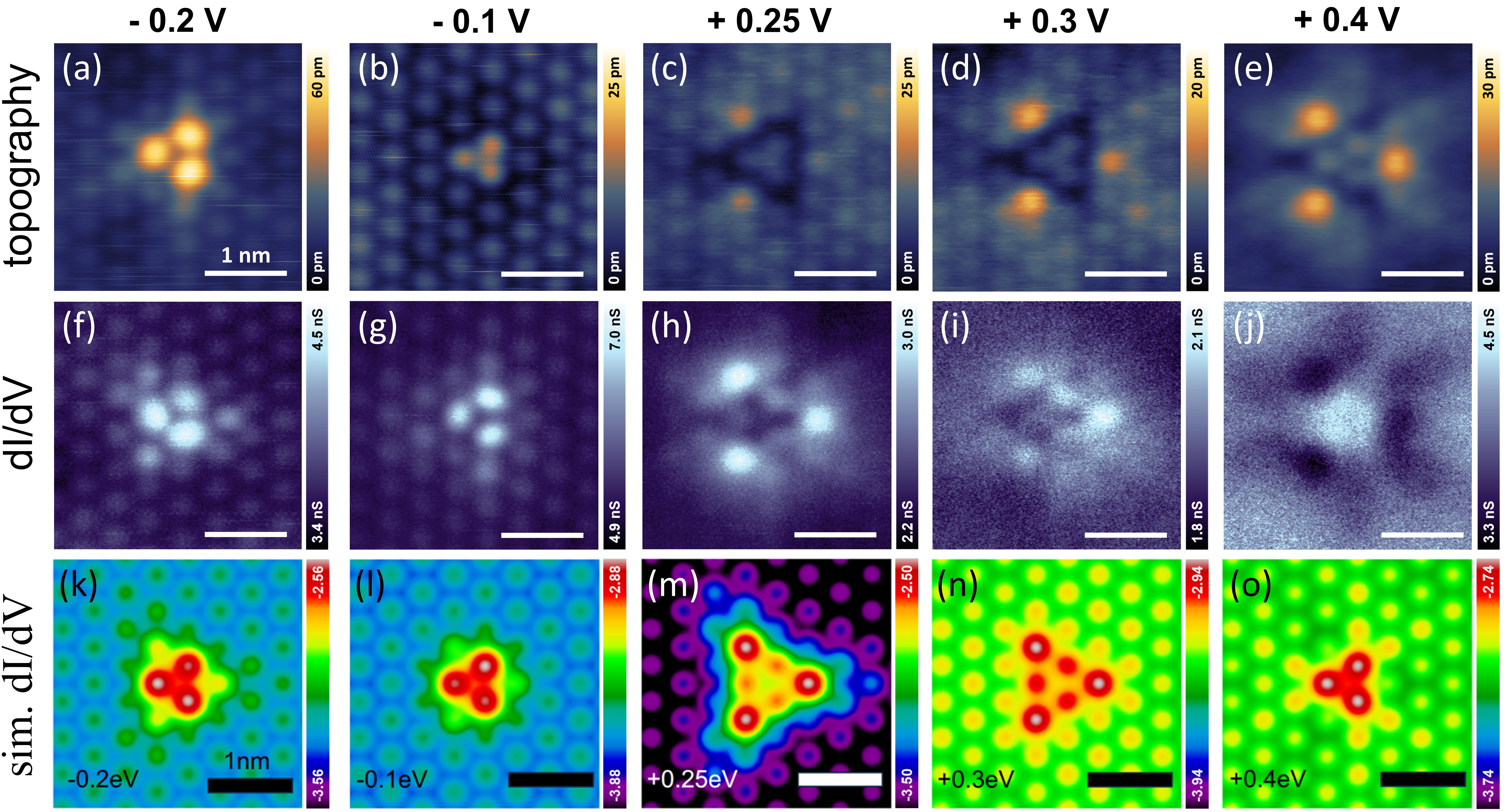}
\caption{Bias-dependent imaging of the fcc-site Fe adatom on Bi$_2$Te$_3$.
(a)--(e)~Constant-current topographic images and
(f)--(j)~Simultaneously acquired $dI/dV$ maps at sample-bias voltages of $-0.2$, $-0.1$, $+0.25$, $+0.3$, and $+0.4$~V, respectively ($I_{\text{set}} = 0.3$~nA, $V_{\text{mod}}=25$~mV).
(k)--(o)~Simulated $dI/dV$ maps from tight-binding calculations at the corresponding energies. The color scale represents $log$(LDOS).}
\label{fig5}
\end{figure*}

In topographic images of fcc-Fe (Fig.~\ref{fig2}(a)), the apparent heights of the three lobes at the nn-Te and, separately, at the nnn-Te sites appear nearly identical, as indicated by the line profile in Fig.~\ref{fig2}(c). A priori, this would be expected from an underlying threefold C$_{3v}$ rotational symmetry of the hollow site. However, the $dI/dV$ image in Fig.~\ref{fig2}(b) reveals a reduced symmetry, evident as distinct contrast levels on the nn-Te sites and in the associated line profile (Fig.~\ref{fig2}(c)). We conclude that the reduced symmetry could not be due to an asymmetric tip apex. In addition to Fig.~\ref{fig2}(a), native defects such as Te vacancies (V\textsubscript{Te}) and Bi antisites (Bi\textsubscript{Te})~\cite{netsou2020identifying} are also imaged with the expected threefold symmetry. Included in the Supplement are measurements with multiple tip apex configurations yielding consistent results (see Supplemental Material Fig.~S1~\cite{supplemental_note}). Notably, we observed the same reduced symmetry in $dI/dV$ imaging with non-magnetic W and PtIr tips as well as the magnetic Ni tip used here, indicating that spin-polarized tunneling does not account for the observed asymmetry. If the asymmetry were tip-induced, all protrusions in the scan would share the same distorted shape, independent of their adsorption site. 

To gain further insight into the origins of the reduced symmetry, we performed DFT calculations of fcc-Fe on the Bi$_2$Te$_3$(111) surface. The Fe adatom was initially placed at the high-symmetry C$_{3v}$ site, in which all three Fe--Te distances are equal ($2.621$~\AA). Upon structural relaxation, the calculations converge to a lower-symmetry C$_{1v}$ configuration, energetically favored by $72.5$~meV. The Fe atom sits approximately $0.77$~\AA\ below the top Te layer, which suggests that it may behave more like a bulk interstitial than an adatom. In this distorted geometry (Fig.~\ref{fig3}), one Fe--Te distance elongates to $2.628$~\AA\ while the remaining two contract to $2.608$~\AA. Although these changes are small ($< 0.5$~\%), they are sufficient to break the threefold equivalence of the nn-Te sites. 

The impact of this JT distortion on the LDOS can be revealed by comparing experimental $dI/dV$ spectroscopy with DFT LDOS calculations. Figure~\ref{fig4}(a) shows $dI/dV$ spectra, taken with the tip positioned at the three nn-Te and three nnn-Te sites, as indicated by the colored markers in the inset. The spectra exhibit three common features: a dip at $-0.2$~V and peaks at $-0.1$~V and $+0.3$~V. These features are absent in the reference Bi$_2$Te$_3$ spectrum (gray curve), and are thus associated with the fcc-Fe. Although these features occur at similar voltages at all tip positions, the spectral weight varies among the three nominally equivalent Te sites. For example, one of the nnn-Te sites (d; cyan) has a noticeably higher signal at the $+0.3$~V peak compared to the otherwise equivalent sites (e, f; blue, violet). Further distinctions between the e and f sites are evident at $-0.1$~V. Similarly, the nn-Te site (c; orange) has distinctly lower $dI/dV$ signals at both $-0.1$ V and $+0.3$ V peaks compared to sites (a, b; green, yellow).

Figure~\ref{fig4}(b) shows the DFT-calculated projected density of states (pDOS) for the relaxed fcc-site Fe adatom. The pDOS is obtained by projecting the Kohn--Sham wavefunctions onto the atom-centered Fe $3d$ orbitals, allowing the orbital-resolved electronic states to be analyzed as a function of energy. The Fe-site pDOS (upper panel) reflects the crystal-field splitting of the Fe $d$ manifold into two dominant groups around the Fermi level. In the occupied states, a pronounced peak is found near $-0.3$~eV, dominated by the Fe $d_{xz}$ and $d_{yz}$ orbitals with smaller contributions from the $d_{xy}$ and $d_{x^2-y^2}$ orbitals. In the unoccupied states, overlapping peaks from $+0.2$--$0.4$~eV predominantly arise from the in-plane $d_{x^2-y^2}$ and $d_{xy}$ orbitals together with the out-of-plane $d_{z^2}$ orbital.

The Fe $d$-orbital weight projected at the six neighboring Te sites (lower panel) shows corresponding peaks at $\sim -0.3$~eV and $\sim +0.3$~eV, indicating that Fe $d$-orbital character extends to the neighboring Te sites through $d$--$p$ hybridization. Both the nn-Te and nnn-Te projections show distinct LDOS on one of the sites compared to the other two. For example, the nnn-Te site (d; cyan) has lower occupied LDOS than the other two sites (e, f; blue, violet). Likewise, nn-Te sites (b, c; yellow, orange) have a slightly higher LDOS in unoccupied states ($>0.2$~eV) than the remaining nn-Te site (a; green). Traces b, c (nn-Te) and traces e, f (nnn-Te) overlap, consistent with the $C_{1v}$ mirror symmetry. The clear orbital selectivity and energetic separation of these features are consistent with the experimentally observed $dI/dV$ peaks at $-0.1$~V and $+0.3$~V. The differing orbital character naturally explains the spatial anisotropy observed in the STM measurements, as the d-orbitals hybridize differently with the triangular Te lattice. 

To summarize these comparisons, Fig.~\ref{fig5} shows STM topographic and $dI/dV$ images at multiple bias voltages around the Fermi level, along with simulated $dI/dV$ maps from our tight-binding calculations. The $dI/dV$ maps reveal a continuous evolution of spatial contrast that directly reflects the orbital character of the impurity states. Notably, neither the experimental nor the simulated $dI/dV$ maps exhibit appreciable contrast at the Fe site for most energies. The imaged LDOS primarily reflects the hybridization of the Fe $d$ orbitals with the neighboring Te $p$ orbitals, likely enhanced by the Fe position below the surface layer. For instance, the empty-state topography at $+0.4$~V (Fig.~\ref{fig5}(e), Fig.~\ref{fig1}(a)) shows that the fcc-site Fe contrast extends several nanometers beyond the nnn-Te sites. The reduced symmetry is observed across all imaging biases, but is more prominent in the filled states. At $-0.2$~V, one of the three nn-Te sites appears relatively dim compared to the other two sites, but this pattern inverts at $-0.1$~V. The inversion of the asymmetric contrast over this narrow energy range is consistent with the two components of the split, predominantly $d_{xz}$ and $d_{yz}$-derived manifold carrying different spatial weights, as reproduced in the simulated $dI/dV$ maps (Fig.~\ref{fig5}(k)--(o)). At positive bias, the spectral weight shifts to the nnn-Te sites, with the highest contrast at $+0.25$~V, which is also reproduced in the simulation (Fig.~\ref{fig5}(m)). The clearest asymmetry, though weak in signal, occurs at $+0.3$~V, where one nnn-Te site appears brighter than the other two. Our tight-binding simulations attribute the dominant nnn-Te contrast at this energy to the Fe $d_{z^2}$ orbital, which hybridizes with both the nn-Te and nnn-Te atoms, suppressing contrast at the nn-Te sites while enhancing it at the nnn-Te sites (Fig.~\ref{fig5}(m,n)). At $+0.4$--$0.5$~V, the $dI/dV$ contrast shifts markedly toward the nn-Te sites (Fig.~\ref{fig5}(j)), signaling a crossover to a different orbital character. This is consistent with the onset of $d_{x^2-y^2}$ and $d_{xy}$ states, which the pDOS calculations place at slightly higher energies than $d_{z^2}$. Lastly, we note that STM topography reflects the integrated density of states from the Fermi level to the bias chosen, and is in qualitative agreement with the evolution in the $dI/dV$ maps with voltage.

The filling of the Fe $3d$ manifold provides the electronic driving force for the Jahn--Teller distortion. Our calculations identify a $3d^6$ ground-state configuration for the fcc-site Fe adatom, consistent with previous x-ray absorption spectroscopy and XMCD measurements of Fe on Bi$_2$Te$_3$~\cite{eelbo_strong_2014}. In the ideal $C_{3v}$ ligand field, the Fe $d$ manifold separates into an $A_1(d_{z^2})$ singlet and two $E$ doublets, $E(d_{xz},d_{yz})$ and $E(d_{xy},d_{x^2-y^2})$. The six $3d$ electrons are distributed among these crystal-field levels, with the occupied states closest to the Fermi level predominantly derived from the partially occupied $E(d_{xz},d_{yz})$ manifold. Upon distortion to $C_{1v}$, the Fe $3d$ occupation is redistributed, with the dominant change being a redistribution of minority-spin weight from the $E(d_{xz},d_{yz})$ manifold toward the $A_1(d_{z^2})$ state. This redistribution is accompanied by a lowering of the total energy by $72.5$~meV, consistent with a Jahn--Teller distortion.

Beyond the local symmetry breaking, we find that the orientation of the three-lobe impurity pattern is not random. In many regions, a large fraction of fcc-site Fe adatoms relax in the same bright-lobe configuration, such that the major axis of the triangular contrast aligns along a single crystallographic direction (see Supplemental Material Fig.~S2~\cite{supplemental_note}). This may be due to residual strain in the Bi$_2$Te$_3$ crystal. This picture closely relates to the aligned Jahn--Teller domains reported for Re$_\text{Mo}$ in MoS$_2$, where epitaxial strain was proposed to set a common relaxation direction across large areas rather than direct dopant--dopant interaction \cite{xiang_charge_2024}. 

In summary, our findings demonstrate that the commonly assumed $C_{3v}$ adsorption geometry does not strictly hold for Fe on Bi$_2$Te$_3$. As shown in Supplemental Material Fig.~S3~\cite{supplemental_note}, we note that hcp-site Fe adatoms also exhibit hints of reduced symmetry in $dI/dV$ imaging, though the effect is less pronounced. Since JT distortions influence magnetic anisotropy~\cite{odkhuu_giant_2016}, which in turn determines whether a mass gap opens at the Dirac point~\cite{sessi_signatures_2014, liu_magnetic_2009, qi_topological_2011}, these structural details have direct consequences for the magnetic and topological properties of TI surfaces. More broadly, extending this picture to other transition-metal impurities and TI platforms suggests that engineering local symmetry breaking--for example through strain, interface design, or doping configuration--may provide a practical pathway for controlling the magnetic anisotropy and band topology of topological materials ~\cite{petrov_intrinsic_2021}.

\section{Acknowledgments}
We thank Roland Kawakami for helpful discussions. This work was supported by the U.S.\ Department of Energy, Office of Basic Energy Sciences, under Award No.\ DE-SC0016379.
\appendix

\section{Tight Binding}
\label{appendixA}
The simulated $dI/dV$ images for the Fe impurity on the fcc-hollow site were calculated by solving Dyson's Equation in real-space~\cite{KosterSlater.PhysRev.1954b,KosterSlater.PhysRev.1954c,KortanFlatte.PRB.2017, Siegl.NanoLett.2025,Sink.PRB.2025,SinkFlatte.2026},
\begin{align}
    \hat{G}(\omega,\mathbf{R},\mathbf{R}')=\left(\hat{1}-\hat{g}(\omega,\mathbf{R},\mathbf{R}')\hat{V}(\mathbf{R},\mathbf{R}')\right)^{-1}\hat{g}(\omega,\mathbf{R},\mathbf{R}')
\end{align}
where $\hat{V}$ is the impurity potential and $\hat{g}$ is the tight-binding homogeneous retarded Green's function solutions for the pristine 3 QL slab, including effects for SOC. The real space Green's functions are computed by taking the numerical inverse Fourier transform of the resolvent of the k-space Hamiltonian,

\begin{align}
    \mathbf{g}(\omega,\mathbf{R},\mathbf{R}')=\int_{BZ}dk^2\left(\omega+i\delta-\hat{H}(\mathbf{k})\right)^{-1}e^{-i\mathbf{k}\cdot(\mathbf{R}-\mathbf{R}')}
\end{align}

where the Hamiltonian is of the form,
\begin{align}
    \hat{H}(\mathbf{k})=\sum_{\alpha,\beta}h_{\alpha,\beta}c^{\dagger}_\alpha c_{\beta} e^{i\mathbf{k}\cdot\mathbf{R}_{\alpha,\beta}}.
\end{align}
The tight-binding parameters for this model were taken from Ref. \cite{kobayashi_electron_2011}. These model parameters for Bi$_2$Te$_3$ were demonstrated to successfully characterize the thickness-dependent formation of the $\Gamma$-centered Dirac cone.

For the Fe impurity potential, we include onsite and hybridization between the Fe \textit{d}-orbitals and the surface Te \textit{p}-orbitals using a Slater-Koster hybridization schema with Harrison distance scaling,
\begin{align}
    \hat{V}=\begin{pmatrix}
        \hat{\epsilon}_{Fe-d}&\hat{V}^\dagger_{pd}\\
        \hat{V}_{pd}&\hat{\epsilon}_{Te-p}
    \end{pmatrix}
\end{align}

where
\begin{align}
\left(\hat{V}_{pd}\right)_{\alpha,\beta}=\Theta_{\alpha,\beta}~pd\sigma\left(\frac{d_0}{d_{\alpha,\beta}}\right)^2+\Phi_{\alpha,\beta}~pd\pi\left(\frac{d_0}{d_{\alpha,\beta}}\right)^2.
\end{align}
The terms $\Theta_{\alpha,\beta}$ and $\Phi_{\alpha,\beta}$ are orbital and orientation dependent prefactors~\cite{slater_koster.PR.1954}.

The Fe onsite \textit{d}-manifold is parameterized by,
\begin{align}
\hat{\epsilon}_{Fe}=(E_d\hat{1}_{l=2}\otimes+\hat{\Delta}_{CF})\sigma_0
\end{align}

where $E_d$ controls the offset of the centroid of the \textit{d}-manifold relative to the Fermi-level and 

\begin{align}
\hat{\Delta}_{CF}=
\{\Delta_{\overline{E}(xy)},\Delta_{E({yz})},\Delta_{E({zx})},\Delta_{\overline{E}({x^2-y^2})},\Delta_{A({z^2})}\}
\end{align}
is crystal field splitting term with the symmetry of the ideal $C_{3v}$ configuration.

The perturbation to the Te onsite \textit{p}-manifold is,
\begin{align}
    \hat{\epsilon}_{Te}=(\delta E_p\hat{1}_{l=1}\otimes\sigma_0)
\end{align}
where $\delta E_p$ is the relative offset of the \textit{p}-orbitals on the three nn-Te atoms that make up the fcc hollow-site.
\bibliography{bibliography}

@PREAMBLE{
 "\providecommand{\noopsort}[1]{}" 
 # "\providecommand{\singleletter}[1]{#1}%" 
}

@Article{Necas2012,
  author = {Ne\v{c}as, David and Klapetek, Petr},
  title = {Gwyddion: an open-source software for {SPM} data analysis},
  journal = {Cent. Eur. J. Phys.},
  pages = {181--188},
  volume = {10},
  number = {1},
  year = {2012},
  doi = {10.2478/s11534-011-0096-2},
}

@article{eelbo_strong_2014,
  title = {Strong out-of-plane magnetic anisotropy of {Fe} adatoms on {Bi}$_{2}${Te}$_{3}$},
  volume = {89},
  doi = {10.1103/PhysRevB.89.104424},
  number = {10},
  journal = {Phys. Rev. B},
  author = {Eelbo, T. and Wa\'{s}niowska, M. and Sikora, M. and Dobrza\'{n}ski, M. and Koz{\l}owski, A. and Pulkin, A. and Aut\`{e}s, G. and Miotkowski, I. and Yazyev, O. V. and Wiesendanger, R.},
  year = {2014},
  pages = {104424},
}

@article{qi_topological_2008,
  title = {{Topological field theory of time-reversal invariant insulators}},
  volume = {78},
  doi = {10.1103/PhysRevB.78.195424},
  number = {19},
  journal = {Phys. Rev. B},
  author = {Qi, Xiao-Liang and Hughes, Taylor L. and Zhang, Shou-Cheng},
  year = {2008},
  pages = {195424},
}

@article{slater_koster.PR.1954,
  title = {Simplified {LCAO} Method for the Periodic Potential Problem},
  author = {Slater, J. C. and Koster, G. F.},
  journal = {Phys. Rev.},
  volume = {94},
  issue = {6},
  pages = {1498--1524},
  numpages = {0},
  year = {1954},
  month = {Jun},
  publisher = {American Physical Society},
  doi = {10.1103/PhysRev.94.1498},
  url = {https://link.aps.org/doi/10.1103/PhysRev.94.1498}
}

@article{Siegl.NanoLett.2025,
author = {Siegl, Manuel and Zanon, Julian and Sink, Joseph and Rodrigues da Cruz, Adonai and Hedgeland, Holly and Curson, Neil J. and Flatt{\'e}, Michael E. and Schofield, Steven R.},
title = {Imaging the Acceptor Wave Function Anisotropy in Silicon},
journal = {Nano Lett.},
volume = {25},
number = {38},
pages = {13996-14001},
year = {2025},
doi = {10.1021/acs.nanolett.5c02675},
URL = {https://doi.org/10.1021/acs.nanolett.5c02675},
}

@article{KortanFlatte.PRB.2017,
  title = {Spatially resolved electronic structure of an isovalent nitrogen center in {GaAs}},
  author = {Plantenga, R. C. and Kortan, V. R. and Kaizu, T. and Harada, Y. and Kita, T. and Flatt\'e, M. E. and Koenraad, P. M.},
  journal = {Phys. Rev. B},
  volume = {96},
  issue = {15},
  pages = {155210},
  year = {2017},
  month = {Oct},
  publisher = {American Physical Society},
  doi = {10.1103/PhysRevB.96.155210}
}

@article{SinkFlatte.2026,
  author = {Joseph Sink and Michael E. Flatté},
  title = {Nuclear-Electron Hyperfine Coupling of the Shallow States Associated with Vacancies in Gallium Nitride},
  journal       = {arXiv preprint},
  archivePrefix = {arXiv},
  eprint        = {2605.26053},
  year          = {2026}
}

@article{KosterSlater.PhysRev.1954b,
  title = {Wave Functions for Impurity Levels},
  author = {Koster, G. F. and Slater, J. C.},
  journal = {Phys. Rev.},
  volume = {95},
  issue = {5},
  pages = {1167--1176},
  year = {1954},
  month = {Sep},
  publisher = {American Physical Society},
  doi = {10.1103/PhysRev.95.1167}
}

@article{KosterSlater.PhysRev.1954c,
  title = {Simplified Impurity Calculation},
  author = {Koster, G. F. and Slater, J. C.},
  journal = {Phys. Rev.},
  volume = {96},
  issue = {5},
  pages = {1208--1223},
  year = {1954},
  month = {Dec},
  publisher = {American Physical Society},
  doi = {10.1103/PhysRev.96.1208}
}

@article{Sink.PRB.2025,
  title = {{Microscopic scattering approach to in-gap states: Cr adatoms on superconducting $\ensuremath{\beta}\text{\ensuremath{-}}{\mathrm{Bi}}_{2}\mathrm{Pd}$}},
  author = {Sink, Joseph and Paudyal, Hari and Jyoti, Divya and Heinrich, Andreas J. and Choi, Deung-Jang and Lorente, Nicol\'as and Flatt\'e, Michael E.},
  journal = {Phys. Rev. B},
  volume = {112},
  issue = {24},
  pages = {245429},
  numpages = {9},
  year = {2025},
  month = {Dec},
  publisher = {American Physical Society},
  doi = {10.1103/75p3-2x2h},
  url = {https://link.aps.org/doi/10.1103/75p3-2x2h}
}

@article{liu_magnetic_2009,
  title = {Magnetic Impurities on the Surface of a Topological Insulator},
  volume = {102},
  doi = {10.1103/PhysRevLett.102.156603},
  number = {15},
  journal = {Phys. Rev. Lett.},
  author = {Liu, Qin and Liu, Chao-Xing and Xu, Cenke and Qi, Xiao-Liang and Zhang, Shou-Cheng},
  year = {2009},
  pages = {156603},
}

@article{sessi_signatures_2014,
  title = {Signatures of {Dirac} fermion-mediated magnetic order},
  volume = {5},
  doi = {10.1038/ncomms6349},
  number = {1},
  journal = {Nat. Commun.},
  author = {Sessi, Paolo and Reis, Felix and Bathon, Thomas and Kokh, Konstantin A. and Tereshchenko, Oleg E. and Bode, Matthias},
  year = {2014},
  pages = {5349},
}

@article{west_identification_2012,
  title = {Identification of magnetic dopants on the surfaces of topological insulators: {Experiment} and theory for {Fe} on $\text{Bi}_2\text{Te}_3(111)$},
  volume = {85},
  doi = {10.1103/PhysRevB.85.081305},
  number = {8},
  journal = {Phys. Rev. B},
  author = {West, D. and Sun, Y. Y. and Zhang, S. B. and Zhang, T. and Ma, Xucun and Cheng, P. and Zhang, Y. Y. and Chen, X. and Jia, J. F. and Xue, Q. K.},
  year = {2012},
  pages = {081305(R)},
}

@article{chen_massive_2010,
  title = {Massive {Dirac} Fermion on the Surface of a Magnetically Doped Topological Insulator},
  volume = {329},
  doi = {10.1126/science.1189924},
  number = {5992},
  journal = {Science},
  author = {Chen, Y. L. and Chu, J.-H. and Analytis, J. G. and Liu, Z. K. and Igarashi, K. and Kuo, H.-H. and Qi, X. L. and Mo, S. K. and Moore, R. G. and Lu, D. H. and Hashimoto, M. and Sasagawa, T. and Zhang, S. C. and Fisher, I. R. and Hussain, Z. and Shen, Z. X.},
  year = {2010},
  pages = {659--662},
}

@article{laref_induced_2020,
  title = {Induced spin textures at $3d$ transition metal--topological insulator interfaces},
  volume = {101},
  doi = {10.1103/PhysRevB.101.220410},
  number = {22},
  journal = {Phys. Rev. B},
  author = {Laref, Slimane and Ghosh, Sumit and Tsymbal, Evgeny Y. and Manchon, Aurelien},
  year = {2020},
  pages = {220410(R)},
}

@article{lee_imaging_2015,
  title = {Imaging {Dirac}-mass disorder from magnetic dopant atoms in the ferromagnetic topological insulator $\text{Cr}_x(\text{Bi}_{0.1}\text{Sb}_{0.9})_{2-x}\text{Te}_3$},
  volume = {112},
  doi = {10.1073/pnas.1424322112},
  number = {5},
  journal = {Proc. Natl. Acad. Sci. U.S.A.},
  author = {Lee, Inhee and Kim, Chung Koo and Lee, Jinho and Billinge, Simon J. L. and Zhong, Ruidan and Schneeloch, John A. and Liu, Tiansheng and Valla, Tonica and Tranquada, John M. and Gu, Genda and Davis, J. C. S\'{e}amus},
  year = {2015},
  pages = {1316--1321},
}

@article{furthmuller1996dimer,
  title = {{Dimer reconstruction and electronic surface states on clean and hydrogenated diamond (100) surfaces}},
  author = {Furthm{\"u}ller, J. and Hafner, J. and Kresse, G.},
  journal = {Phys. Rev. B},
  volume = {53},
  number = {11},
  pages = {7334},
  year = {1996},
  doi = {10.1103/PhysRevB.53.7334},
}

@article{kresse1996efficiency,
  title = {{Efficiency of ab-initio total energy calculations for metals and semiconductors using a plane-wave basis set}},
  author = {Kresse, G. and Furthm{\"u}ller, J.},
  journal = {Comput. Mater. Sci.},
  volume = {6},
  number = {1},
  pages = {15--50},
  year = {1996},
  doi = {10.1016/0927-0256(96)00008-0},
}

@article{blochl1994projector,
  title = {{Projector augmented-wave method}},
  author = {Bl{\"o}chl, P. E.},
  journal = {Phys. Rev. B},
  volume = {50},
  number = {24},
  pages = {17953},
  year = {1994},
  doi = {10.1103/PhysRevB.50.17953},
}

@article{perdew1996o,
  title = {Generalized Gradient Approximation Made Simple},
  author = {Perdew, J. P. and Burke, K. and Ernzerhof, M.},
  journal = {Phys. Rev. Lett.},
  volume = {77},
  number = {18},
  pages = {3865--3868},
  year = {1996},
  doi = {10.1103/PhysRevLett.77.3865},
}

@article{xiang_charge_2024,
  title = {Charge state-dependent symmetry breaking of atomic defects in transition metal dichalcogenides},
  volume = {15},
  doi = {10.1038/s41467-024-47039-4},
  number = {1},
  journal = {Nat. Commun.},
  author = {Xiang, Feifei and Huberich, Lysander and Vargas, Preston A. and Torsi, Riccardo and Allerbeck, Jonas and Tan, Anne Marie Z. and Dong, Chengye and Ruffieux, Pascal and Fasel, Roman and Gr{\"o}ning, Oliver and Lin, Yu-Chuan and Hennig, Richard G. and Robinson, Joshua A. and Schuler, Bruno},
  year = {2024},
  pages = {2738},
}

@article{odkhuu_giant_2016,
  title = {Giant perpendicular magnetic anisotropy of an individual atom on two-dimensional transition metal dichalcogenides},
  volume = {94},
  doi = {10.1103/PhysRevB.94.060403},
  number = {6},
  journal = {Phys. Rev. B},
  author = {Odkhuu, Dorj},
  year = {2016},
  pages = {060403(R)},
}

@article{kobayashi_electron_2011,
  title = {Electron transmission through atomic steps of {Bi}$_{2}${Se}$_{3}$ and {Bi}$_{2}${Te}$_{3}$ surfaces},
  volume = {84},
  doi = {10.1103/PhysRevB.84.205424},
  number = {20},
  journal = {Phys. Rev. B},
  author = {Kobayashi, Katsuyoshi},
  year = {2011},
  pages = {205424},
}

@article{petrov_intrinsic_2021,
  title = {Intrinsic Magnetic Topological Insulator State Induced by the {Jahn}--{Teller} Effect},
  volume = {12},
  doi = {10.1021/acs.jpclett.1c02396},
  number = {37},
  journal = {J. Phys. Chem. Lett.},
  author = {Petrov, E. K. and Ernst, A. and Menshchikova, T. V. and Chulkov, E. V.},
  year = {2021},
  pages = {9076--9085},
}

@article{chang_experimental_2013,
  title = {Experimental Observation of the Quantum Anomalous Hall Effect in a Magnetic Topological Insulator},
  volume = {340},
  doi = {10.1126/science.1234414},
  number = {6129},
  journal = {Science},
  author = {Chang, Cui-Zu and Zhang, Jinsong and Feng, Xiao and Shen, Jie and Zhang, Zuocheng and Guo, Minghua and Li, Kang and Ou, Yunbo and Wei, Pang and Wang, Li-Li and Ji, Zhong-Qing and Feng, Yang and Ji, Shuaihua and Chen, Xi and Jia, Jinfeng and Dai, Xi and Fang, Zhong and Zhang, Shou-Cheng and He, Ke and Wang, Yayu and Lu, Li and Ma, Xu-Cun and Xue, Qi-Kun},
  year = {2013},
  pages = {167--170},
}

@article{qi_topological_2011,
  title = {Topological insulators and superconductors},
  volume = {83},
  doi = {10.1103/RevModPhys.83.1057},
  number = {4},
  journal = {Rev. Mod. Phys.},
  author = {Qi, Xiao-Liang and Zhang, Shou-Cheng},
  year = {2011},
  pages = {1057--1110},
}

@article{netsou2020identifying,
  title={Identifying Native Point Defects in the Topological Insulator {Bi$_2$Te$_3$}},
  author={Netsou, Asteriona-Maria and Muzychenko, Dmitry A. and Dausy, Heleen and Chen, Taishi and Song, Fengqi and Schouteden, Koen and Van Bael, Margriet J. and Van Haesendonck, Chris},
  journal={ACS Nano},
  volume={14},
  number={10},
  pages={13172--13179},
  year={2020},  doi={10.1021/acsnano.0c04861}
}

@misc{supplemental_note,
  note = {See Supplemental Material at [URL will be inserted by publisher] for tip characterization, directional-domain analysis, and additional data.},
}
\end{document}